\documentclass{revtex4}

\usepackage{graphicx}
\usepackage{bm}
\usepackage{amssymb}
\usepackage{amsmath}
\usepackage{float}

\usepackage[english]{babel}
\usepackage[latin1]{inputenc}

\begin{document}

\title{Upper bounds for critical coupling constants for binding some quantum many-body systems}

\author{Clara \surname{Tourbez}}
\email[E-mail: ]{clara.tourbez@student.umons.ac.be}
\thanks{ORCiD: 0009-0004-0909-6974}
\author{Claude \surname{Semay}}
\email[E-mail: ]{claude.semay@umons.ac.be}
\thanks{ORCiD: 0000-0001-6841-9850}
\author{Cyrille \surname{Chevalier}}
\email[E-mail: ]{cyrille.chevalier@umons.ac.be}
\thanks{ORCiD: 0000-0002-4509-4309}
\affiliation{Service de Physique Nucl\'{e}aire et Subnucl\'{e}aire,
Universit\'{e} de Mons,
UMONS Research Institute for Complex Systems,
Place du Parc 20, 7000 Mons, Belgium}
\date{\today}

\begin{abstract}
When particles interact via two-body short-range central potential wells, binding can occur for some critical values of the coupling constants. Using the envelope theory, upper bounds  for critical coupling constants are computed for quantum nonrelativistic systems containing identical particles and systems containing identical particles plus a different one. 
\end{abstract}

\maketitle

\section{Introduction}
\label{sec:intro}

In nonrelativistic quantum many-body systems, binding can occur if some conditions are fulfilled. The potentials must be sufficiently attractive and the particle masses sufficiently large \cite{rich94,goy95,mosz00,rich20}. In this paper, we consider only the case of two-body central potential wells with the structure 
\begin{equation}
\label{Vgv}
V_{ij}(|\bm{r}_i-\bm{r}_j|) = - g_{ij}\, v_{ij}(|\bm{r}_i-\bm{r}_j|).
\end{equation}
In this equation, $g_{ij}$ is a coupling constant with the dimension of an energy and $v_{ij}(r)$ a dimensionless short-range ``globally" positive function. This implies  that $dv_{ij}(r)/dr$ is a short-range ``globally" negative function. At least some coupling constants must be positive to insure the existence of bound states. If a $N$-body system depends only on a single coupling constant $g$, its critical coupling constant $g_N(\{q_\alpha\})$ is defined as the value at which a bound state with a given set of quantum numbers $\{q_\alpha\}$ appears for the $N$ particles. This bound state exists for any $g\ge g_N(\{q_\alpha\})$, the equality corresponding to a state with a zero energy. In the following, the mention of the quantum numbers will be omitted. The definition is similar if several coupling constants are present. The study of critical coupling constants can be useful in nuclear and atomic physics in which systems with $N$ particles are bound whereas systems with $N-1$ particles are not bound \cite{naid17}. Applications in materials science could also be considered since it is possible to modulate the forces between atoms \cite{han24}.

The computation of critical coupling constants is very challenging, specially for many-body systems. Indeed, systems with a very low binding energy or even a vanishing energy have a wavefunction very extended in space. So it could be difficult to guarantee the accuracy of numerical calculations that can turn out to be very long. Obtaining reliable information about these critical coupling constants is then interesting. This is the purpose of this paper.

Two types of many-body systems are considered: a first one containing identical particles, treated in Sec.~\ref{sec:n}, and a second one containing identical particles plus a different one, treated in Sec.~\ref{sec:npo}. Upper bounds for the corresponding critical coupling constants are computed with the envelope theory (ET) \cite{hall80,hall83,hall04,silv10,sema13a,sema23}. The ET is a technique to compute approximate solutions of many-body systems with arbitrary kinematics in $D$ dimensions. The basic idea is to replace the Hamiltonian $H$ considered by a many-body harmonic oscillator Hamiltonian $\tilde H$ which is solvable \cite{cint21}. $\tilde H$ depends on parameters which are optimized in such way that one of its particular eigenvalues is as close as possible to the corresponding eigenvalue of $H$. The main interest of this method is that the computational cost is independent from the number of particles. Moreover, the approximate eigenvalues can be analytical upper or lower bounds in some favourable situations. This is used in this work to obtain upper bounds of critical coupling constants for the many-body systems mentioned above. The case of all particles identical has already been studied elsewhere \cite{silv10,sema13a}, but it is developed here for completeness and also because new results are presented. Especially, checks for the variational character of the critical coupling constants are performed for two systems: $N=2$ with $D=3$ and $3\le N \le 100$ with $D=1$. A brief summary is given in Sec.~\ref{sec:sum}.
 
\section{Many identical particles}
\label{sec:n}

Let us start by giving the properties of the ET for $N$ identical particles in $D$ dimensions with pairwise central forces. The generic Hamiltonian is written
\begin{equation}
\label{HNid}
H=\sum_{i=1}^N T(|\bm p_i|) + \sum_{i<j=2}^N V(|\bm{r}_i-\bm{r}_j|),
\end{equation}
where $T$ is the kinetic energy, $V$ is a two-body central potential, and where the centre of mass motion is removed ($\sum_{i=1}^N \bm p_i = \bm 0$). An ET approximate eigenvalue $E$ for a completely (anti)symmetrised state is given by the following set of equations \cite{sema13a,chev22}
\begin{align}
\label{ETNid1}
&E=N\, T(p_0) + C^2_N\, V (\rho_0), \\
\label{ETNid3}
&N\, p_0\,T'(p_0)  =  C^2_N\, \rho_0\, V'(\rho_0), \\
\label{ETNid2}
&Q(N)\hbar=\sqrt{C^2_N}\,\rho_0\, p_0, 
\end{align}
where $U'(x)=dU/dx$, $C^2_N=N(N-1)/2$ is the number of pairs, $\rho_0$ (linked to the mean distance between two particles) and $p_0$ (linked to the mean momentum of a particle) are two positive parameters to be determined with (\ref{ETNid3})-(\ref{ETNid2}), and 
\begin{equation}
\label{QN}
Q(N) = \sum_{i=1}^{N-1} \left(2\, n_i + l_i + \frac{D}{2}\right) 
\end{equation}
is a global quantum number for $D \ge 2$. For $D=1$, \cite{sema19}
\begin{equation}
\label{QND1}
Q(N) = \sum_{i=1}^{N-1} \left(n_i + \frac{1}{2}\right).
\end{equation}
The quantum numbers $\{ n_i,l_i\}$ are associated with the $N-1$ internal variables. 
The allowed values of $Q(N)$ depend on the bosonic/fermionic nature of the particles \cite{cimi24}. Its minimum value, $(N-1)D/2$, is reached for the bosonic ground state. Reliable eigenvalues can be obtained for a large variety of systems \cite{sema15a,lore23}. In some favourable situations, the ET has a variational character \cite{hall80,sema23,sema20}. It is also possible to obtain relations between energies of many-body systems with different numbers of particles or different masses \cite{silv11}. 

For $D \ge 2$, it is possible to improve the ET by breaking the strong degeneracy inherent to the method (see~(\ref{QN})). The idea is to combine the ET with the dominantly orbital state method \cite{sema13b}. This results in dividing the global quantum number $Q(N)$ into angular and radial contributions by the introduction of a parameter $\phi$ \cite{chev22,sema15b}:
\begin{align}
\label{Qnulambda}
Q_\phi(N) &= \phi\, \nu + \lambda \quad \textrm{where} \\
\label{nulambda}
\nu &= \sum_{i=1}^{N-1} \left (n_i + \frac{1}{2} \right) \quad \textrm{and} \quad
\lambda =\sum_{i=1}^{N-1}\left (l_i + \frac{D-2}{2} \right).
\end{align}
This is inspired from the existence of an effective quantum number that determines with high accuracy the level ordering of centrally symmetric 2-body systems \cite{loba09}. In (\ref{Qnulambda})-(\ref{nulambda}), the same assumption is extended to $N > 2$. Note that the original quantum number $Q(N)$ is recovered with $\phi = 2$. The procedure consists of four steps:
\begin{enumerate}
  \item Choose the values $\nu$ and $\lambda$ for the chosen $Q(N)$.
  \item Find the values $\tilde \rho_0$ and $\tilde p_0$ by solving (\ref{ETNid3})-(\ref{ETNid2}) but with $Q(N)$ replaced by $\lambda$.
  \item Compute the value of $Q_\phi(N)$ with $\phi$ given by
\begin{equation}
\label{phi}
\phi = \left[ 2 + \frac{\tilde p_0\, T''(\tilde p_0)}{T'(\tilde p_0)} + \frac{\tilde \rho_0\, V''(\tilde \rho_0)}{V'(\tilde \rho_0)} \right]^{1/2}.
\end{equation}
  \item Solve (\ref{ETNid1})-(\ref{ETNid2}) but with $Q(N)$ replaced by $Q_\phi(N)$.
\end{enumerate}
The eigenvalues can be remarkably improved for systems with all particles identical \cite{sema15b}. A drawback is that the possible variational character is no longer guaranteed. Equation~(\ref{phi}) is the same as equation~(13) in \cite{chev22} after the implementation of unnoticed simplifications giving a more elegant and symmetrical formula. In the nonrelativistic case, (\ref{phi}) reduces to
\begin{equation}
\label{phinr}
\phi_{\textrm{NR}} = \left[ 3 + \frac{\tilde \rho_0\, V''(\tilde \rho_0)}{V'(\tilde \rho_0)} \right]^{1/2}.
\end{equation}
Moreover, if $V(r) \propto r^q$, then $\phi_{\textrm{NR}}=\sqrt{2+q}$. This yields the exact spectra for $q=2$ with arbitrary values of $N$, and for $q=-1$ with $N=2$ \cite{sema15a}. The structure of $\phi_{\textrm{NR}}$ is more or less complicated depending on the structure of the potential (see the example below).

As we focus on nonrelativistic many-body quantum systems with short-range central pairwise forces, the generic Hamiltonian of the first type of many-body systems under study is 
\begin{equation}
\label{HNid}
H=\sum_{i=1}^N \frac{\bm p_i^2}{2 m} -\sum_{i<j=2}^N g\,v(|\bm{r}_i-\bm{r}_j|).
\end{equation}
With $T(p)=p^2/(2 m)$ and $V(r)=-g\,v(r)$, the critical coupling constant $g_N$ for $N$ identical particles can be computed by setting $E=0$ in (\ref{ETNid1}) for a fixed value of $Q(N)$ associated with a given set of quantum numbers $\{n_i,l_i\}$ \cite{silv10,sema13a}. This yields
\begin{align}
\label{g}
&g_N= \frac{1}{\rho_0^2\, v(\rho_0)}\frac{2}{N(N-1)^2}\frac{Q(N)^2\hbar^2}{m} \quad \textrm{with} \\
\label{rho0}
&2\, v(\rho_0) + \rho_0\, v'(\rho_0)=0.
\end{align}
The intermediate positive quantity $\rho_0$ is fixed by (\ref{rho0}) and used in (\ref{g}). A solution is always possible since $v(\rho_0)$ is expected to be positive and $v'(\rho_0)$ is expected to be negative with the short-range potentials considered. This $N$-body system can only be bound if $g > 0$. This is in agreement with (\ref{g}) which predicts a positive value for $g_N$, since $v(\rho_0)$ is expected to be positive. Values of $g > g_N$ correspond then to bound states with the same set $\{n_i,l_i\}$ within the ET. Note that similar formulas exist for one-body forces \cite{sema13a}, cyclic systems \cite{sema17}, and systems with a special type of many-body forces \cite{sema18}. For the bosonic ground state, (\ref{g}) implies that
\begin{equation}
\label{gnonp1}
\frac{g_N}{g_{N-1}} = \frac{N-1}{N}, 
\end{equation}
which is compatible with results found in \cite{rich94,goy95}. If we assume that the dimensionless potential $v(r)$ depends only on one inverse length $\mu$, dimensional analysis implies that 
\begin{equation}
\label{gdim}
g_N \propto \frac{\mu^2\hbar^2}{m}, 
\end{equation}
which is in agreement with (\ref{g}). The contribution from the structure of the potential well is factorised in the quantity $(\rho_0^2\, v(\rho_0))^{-1}$, the parameter $\rho_0$ being defined by~(\ref{rho0}). 

For nonrelativistic systems, one can define $b(x)$ such that $b(x^2)= V(x)$. If $d^2b(x)/dx^2$ is a concave (convex) function for all positive values of $x$, an approximate ET energy is an upper (lower) bound of the genuine energy \cite{sema23,sema20}. If it is not the case, the variational character is not guaranteed. It is easy to show that if $E$ is an upper (lower) bound of the energy, $g_N$ is then an upper (lower) bound of the genuine critical coupling constant. Actually, the ET relies on the possibility to envelop the potential $V(r)$ with a family of well-chosen harmonic potentials. The variational character of the method is guaranteed when all these harmonic potentials are tangent at a single point of $V(r)$ but do not cross $V(r)$ \cite{sema23}. In this paper, $V(r)$ is a short-range potential well. So, crossing are avoided only for harmonic potentials all above $V(r)$. Only upper bounds of the energies, and of the critical coupling constants, can be obtained in this situation. 

Upper bounds for the eigenvalues and the critical coupling constants are provided by the ET for the three different positive monotonous functions $v(r)$ in Table~\ref{tab:vfac}. The corresponding values of $(\rho_0^2\, v(\rho_0))^{-1}$ are very similar, as expected. Indeed, for very low binding energies, the wave function has a very large extension and it is little influenced by the detailed structure of the short-range potential. Nevertheless, this also shows some limitations of the method since Yukawa and Gaussian potentials yield exactly the same critical coupling constant. 

\begin{table}[H]
\begin{center}
    \begin{tabular}{cc}
\hline\hline
$v(r)$ & $\dfrac{1}{\rho_0^2\, v(\rho_0)}$ \\[8pt]
\hline
$\dfrac{e^{-\mu r}}{\mu r}$ & $e \mu^2$ \\
$e^{-\mu r}$ & $\dfrac{e^2 \mu^2}{4}$ \\
$e^{-\mu^2 r^2}$ & $e \mu^2$ \\
\hline\hline
    \end{tabular} 
    \caption{Values of the factor $(\rho_0^2\, v(\rho_0))^{-1}$ in (\ref{g}) for three dimensionless potential wells $v(r)$. \label{tab:vfac}}
\end{center}
\end{table}

The notion of critical coupling constant is not relevant for a power-law potential since there is no short-range parameter such $\mu$ which can be defined. Nevertheless, if we write, rather artificially, $V(r)=\textrm{sgn}(p)\,g\,(\mu r)^p=-g\, v(r)$ with $p>-2$ and $p\ne 0$, (\ref{g})-(\ref{rho0}) predicts $g_N=0$ for $-2 < p < 0$ and $g_N=\infty$ for $p > 0$. These limit values are the ones expected to produce states with a vanishing energy for attractive power-law potentials. 

Since the introduction of a parameter $\phi$ brings improvements for the ET eigenvalues, it is perhaps possible to take advantage of its existence for the computation of $g_N$. The problem is that $\phi_{\textrm{NR}}$ is a function of the potential by (\ref{phinr}). So, $\phi_{\textrm{NR}}=\phi_{\textrm{NR}}(g)$ depends on $g$, as well as $Q_{\phi_{\textrm{NR}}(g)}(N)$ by (\ref{Qnulambda}). The idea is then to find $g_N$ by solving the transcendental equation 
\begin{equation}
\label{gphi}
g_N= \frac{1}{\rho_0^2\, v(\rho_0)}\frac{2}{N(N-1)^2}\frac{Q_{\phi_{\textrm{NR}}(g_N)}(N)^2\hbar^2}{m} ,
\end{equation}
with $\rho_0$ still defined by (\ref{rho0}). With this formula, the possible variational character of $g_N$ is not guaranteed.

\subsection{Test with an exponential potential}
\label{sec:exp}

In order to put some numbers in these equations, let us first consider $v(r) = e^{-\mu r}$. The ET energy is given by \cite{cimi24}
\begin{equation}
\label{Eexp}
E = -\frac{N(N-1)}{2} \, g \left( 1+ \frac{3}{2} W_0(-Z) \right)e^{3\,W_0(-Z)} \quad \textrm{with} \quad 
Z =\frac{1}{3}\left( \frac{4}{N(N-1)^2} \frac{\mu^2}{g} \frac{Q(N)^2\hbar^2}{m} \right)^{1/3},
\end{equation}
where $W_0$ is the principal branch of the Lambert function \cite{corl96}. The structure of $W_0$ is such that only a finite number of bound states is possible, as expected. The computation of $\phi_{\textrm{NR}}$ yields 
\begin{equation}
\label{phiexp}
\phi_{\textrm{NR}} = \sqrt{3(1+W_0(-\tilde Z))},
\end{equation}
where $\tilde Z$ is the quantity $Z$ in (\ref{Eexp}) computed with $Q(N)$ replaced by $\lambda$. The improved ET energy is then given by (\ref{Eexp}) in which $Q(N)$ is replaced by $Q_{\phi_{\textrm{NR}}}(N)$. It is possible to easily test the quality of the ET approximations for $D=3$ and $N=2$. Without loss of generality, one can set $\hbar=m=\mu=1$, so the Hamiltonian considered is written
\begin{equation}
\label{Hexpwd}
H=\bm p^2 - g\, e^{-|\bm r|}.
\end{equation}
Eigenvalues and critical coupling constants for this Hamiltonian can be computed exactly for $l=0$ \cite{silv09} and accurately computed numerically for $l\ne 0$ with the Lagrange mesh method \cite{baye15}. Table~\ref{tab:exp2} shows that the approximate ET energies are reasonable upper bounds and the improved ones are closer to exact values. In Table~\ref{tab:ccc}, one can see that the critical coupling constants given by (\ref{g})-(\ref{rho0}) are upper bounds and that (\ref{gphi}) gives better values. If the accuracy is not very good, the correct order of magnitude is obtained, specially for (\ref{gphi}). 

\begin{table}[H]
\begin{center}
\caption{Eigenenergies of (\ref{Hexpwd}) with $g=40$, for two identical particles, as a function of $l$ (even for bosons and odd for fermions) and $n$ quantum numbers. For each set of $\{ l\}$,
first line: exact result (see text),
second line: upper bounds from formula~(\ref{Eexp}),
third line: approximations from formula~(\ref{Eexp}) with $Q_{\phi_{\textrm{NR}}}(N)$.
A * indicates that a negative value is not found with the ET. \label{tab:exp2}}
\begin{tabular}{@{}ccccc}
\hline\hline
$l$ & $n=0$ & 1 & 2 & 3 \\
\hline
0 & $-17.5$ & $-6.88$ & $-1.87$ & $-0.08$ \\
  & $-15.7$ & $-3.65$ & * & * \\
  & $-17.3$ & $-5.94$ & $-0.02$ & * \\
1 & $-10.14$ & $-3.35$ & $-0.42$ &  \\
  & $-8.56$ & $-0.37$ & * &  \\
  & $-9.92$ & $-2.41$ & * &  \\
2 & $-5.03$ & $-0.93$ &  &  \\
  & $-3.65$ & * &   &   \\
  & $-4.84$ & * &   &   \\
3 & $-1.55$ &   &   &   \\
  & $-0.37$ &   &   &   \\
  & $-1.36$ &   &   &   \\
\hline\hline
\end{tabular}
\end{center}
\end{table}

\begin{table}[H]
\begin{center}
\caption{Critical coupling constant $g_2$ for some eigenstates of Hamiltonian~(\ref{Hexpwd}) with $\{n,l\}$ quantum numbers for two identical particles ($l$ even for bosons and odd for fermions). For each set of $\{ n\}$,
first line: exact result (see text),
second line: upper bounds from formulas~(\ref{g})-(\ref{rho0}),
third line: approximations from formula~(\ref{gphi}). \label{tab:ccc}}
    \begin{tabular}{cccc}
        \hline\hline
        $n$ & $l=0$ & 1 & 2 \\
        \hline
	0 & 1.45 & 7.05 & 16.3 \\
	  & 4.16 & 11.5 & 22.6 \\
	  & 2.92 & 8.71 & 18.1 \\
	1 & 7.62 & 16.9 & 29.9 \\
	  & 22.6 & 37.4 & 55.9 \\
	  & 16.0 & 26.3 & 40.2 \\
	2 & 18.7 & 31.5 & 48.1 \\
	  & 55.9 & 78.0 & 104. \\
	  & 39.8 & 54.7 & 73.1 \\
        \hline\hline
    \end{tabular} 
\end{center}
\end{table}

\subsection{Test with a Gaussian potential}
\label{sec:gauss}

Let us now consider $v(x) = e^{-\mu^2 x^2}$ for $D=1$. The corresponding Hamiltonian is
\begin{equation}
\label{HNidD1}
H=\sum_{i=1}^N \frac{p_i^2}{2 m} - \sum_{i<j=2}^N g\,e^{-\mu^2(x_i-x_j)^2}.
\end{equation}
The ET upper bound for the bosonic ground state is given by \cite{sema19}
\begin{equation}
\label{Egauss}
E = -\frac{N(N-1)}{2} \, g\, Y^2 \frac{1+2\, W_0(-Y)}{W_0(-Y)^2}   \quad \textrm{with} \quad 
Y =\frac{\mu\,\hbar}{2\sqrt{2 \, m\, g\, N}}.
\end{equation}
Again, the presence of $W_0$ implies that only a finite number of bound states is possible, as expected. As $D=1$, no improvement is possible by using the dominantly orbital state method. The quality of this upper bound with respect to $N$ can be checked thanks to the very accurate results computed in \cite{baye17} for $N=\{3,5,20,100\}$. 

In the following, a unique set of parameters used in \cite{baye17}, $\{ m=1/43.281307,\mu=\hbar=1\}$, is considered for Hamiltonian~(\ref{HNidD1}). Results from Table~\ref{tab:Gauss1} show that the accuracy of the ET bounds improves with the number of particles, that is to say when the binding increases. For a too small number of particles or a weak binding, no reliable ET upper bound can be obtained. This was already noticed for Gaussian interactions with $D=3$ \cite{sema15a}. Upper bounds for the critical coupling constants of the bosonic ground state are given in Table~\ref{tab:Gauss2} as a function of $N$. We have no accurate results to which compare these values, but it is possible to compute reliable estimations of the binding energies for not too small coupling constants (see Sec.~\ref{sec:hyper}). We already know from Table~\ref{tab:Gauss1} that systems with $N\ge 3$ are bound for $g=10/\sqrt{\pi}\approx 5.64$. As $g_N$ is an upper bound of the exact critical coupling constant, it is expected that the $N$-body system is still bound for $g=g_N$. This is the case, as shown in Table~\ref{tab:Gauss2}, since all energies are negative. The energies are still negative and the systems bound for $g=g_N/2$. These numbers give an idea of how far our upper bounds are from the genuine values. This indicate that all values of $g_N$ are more than two times the exact critical coupling constants. No calculation for lower coupling constants is presented because the reliability of low binding energies cannot be guaranteed by our procedure  (see Sec.~\ref{sec:hyper}).  

\begin{table}[H]
\begin{center}
\caption{Bosonic ground state of Hamiltonian~(\ref{HNidD1}). Accurate results from \cite{baye17} are compared with upper bounds~(\ref{Egauss}). A * indicates that a negative value is not found with the ET. The parameters are taken from \cite{baye17}: $m=1/43.281307$, $\mu=\hbar=1$ and $g=10/\sqrt{\pi}$.
\label{tab:Gauss1}}
\begin{tabular}{@{}ccccc}
\hline\hline
$N$ & 3 & 5 & 20 & 100 \\
\hline
\cite{baye17} & $-1.9325$ & $-9.2852$ & $-417.70$ & $-18552$ \\
(\ref{Egauss}) & * & * & $-245$ & $-17553$ \\
\hline\hline
\end{tabular}
\end{center}
\end{table}

\begin{table}[H]
\begin{center}
\caption{Upper bound $g_N$ of the critical coupling constant given by~(\ref{g})-(\ref{rho0}) for the bosonic ground state of Hamiltonian~(\ref{HNidD1}), with binding energies computed with~(\ref{K0})-(\ref{V0K0}) for $g=g_N$ and $g=g_N/2$. The parameters $m=1/43.281307$ and $\mu=\hbar=1$ are taken from \cite{baye17}. 
\label{tab:Gauss2}}
\begin{tabular}{@{}ccccc}
\hline\hline
$N$ & 3 & 5 & 20 & 100 \\
\hline
$g_N$ & $19.6$ & $11.8$ & $2.94$ & $0.588$ \\
$E(g_N)$ & $-14.5$ & $-32.2$ & $-149$ & $-768$ \\
$E(g_N/2)$ & $-4.30$ & $-9.78$ & $-44$ & $-226$ \\
\hline\hline
\end{tabular}
\end{center}
\end{table}

The ET can give reliable upper bounds for the binding energy and good approximations for some observables of various types of many-body systems composed of identical particles \cite{sema19,sema15a}. The two systems considered above show that upper bounds of critical coupling constants can also be computed for this kind of many-body systems in the framework of the ET. Our results are less accurate than the numerical ones obtained in \cite{rich94,goy95,mosz00,rich20} with Gaussian expansions of trial states, but they are very general and can be applied to an arbitrary number of bosons or fermions in dimension $D$ for any state of excitation.

\section{Many identical particles plus a different one}
\label{sec:npo}

With the ET, it is possible to treat many-body systems in dimension $D$ with different particles, especially systems with identical particles plus a different one. For $N_a$ particles $a$ interacting with each other via the potential $V_{aa}$ and interacting with a single particle $b$ via the potential $V_{ab}$, the generic Hamiltonian is written
\begin{equation}
\label{HNp1gen}
H=\sum_{i=1}^{N_a} T_a(|\bm p_i|) + T_b(|\bm p_N|) +\sum_{i<j=2}^{N_a} V_{aa}(|\bm{r}_i-\bm{r}_j|) +\sum_{i=1}^{N_a} V_{ab}(|\bm{r}_i-\bm{r}_N|),
\end{equation}
with $N=N_a+1$. The ET approximations for this Hamiltonian can be computed by solving the following set of equations \cite{cimi22}
\begin{align}
\label{eqn+11}
        &E=N_aT_a\left(p'_a\right)+T_b\left(P_0\right)+C^2_{N_a}V_{aa}\left(r_{aa}\right)+N_aV_{ab}\left(r_0'\right), \\
        &N_aT'_a(p'_a)\frac{p_a^2}{p'_a}=C^2_{N_a}V'_{aa}(r_{aa})r_{aa}+\frac{N_a-1}{2}V'_{ab}(r'_0)\frac{r_{aa}^2}{r'_0}, \\
       &\frac{1}{N_a} T'_a(p'_a)\frac{P_0^2}{p'_a}+T'_b(P_0)P_0=N_aV'_{ab}(r'_0)\frac{R_0^2}{r'_0}, \\
        &Q(N_a)\hbar=\sqrt{C^2_{N_a}}p_a r_{aa}, \\
\label{eqn+15}
        &Q(2)\hbar=P_0 R_0,
\end{align}
where the four positive parameters to be determined in these equations are now $p_a$, $r_{aa}$, $P_0$ and $R_0$, with ${p'_a}^2=p_a^2+\frac{P_0^2}{N_a^2}$ and ${r'_0}^2=\frac{N_a-1}{2N_a}r_{aa}^2+R_0^2$. The situation is clearly much more complex than for all the particles identical. But the ET eigenenergies can be upper or lower bounds in some favourable situations \cite{sema20}. The allowed values of $Q(N_a)$ depend on the bosonic/fermionic nature of the particle $a$, but there is no restriction on the values of $Q(2)$. It is also possible to improve the results from (\ref{eqn+11})-(\ref{eqn+15}) by introducing two parameters ``$\phi$" into the two global quantum numbers $Q(N_a)$ and $Q(2)$, but the procedure is complex and only brings slight improvements, while no longer guaranteeing the possible variational character of the method \cite{chev22}. So, this is not considered in this paper. 

As we focus on nonrelativistic many-body quantum systems with short-range central pairwise forces, the generic Hamiltonian of the second type of many-body systems under study is
\begin{equation}
\label{HNp1}
H=\sum_{i=1}^{N_a} \frac{\bm p_i^2}{2 m_a} + \frac{\bm p_N^2}{2 m_b} -\sum_{i<j=2}^{N_a} g_{aa}\,v_{aa}(|\bm{r}_i-\bm{r}_j|)-\sum_{i=1}^{N_a}g_{ab}\,v_{ab}(|\bm{r}_i-\bm{r}_N|).
\end{equation}
The values for the critical coupling constants $g_{aaN}$ and $g_{abN}$ are also determined by setting $E=0$ in (\ref{eqn+11}). After some tedious calculations, they are given by
\begin{align}
\label{gabn}
g_{abN}v_{ab}'(r_0') &= -\frac{1}{N_a} \frac{r_0'}{R_0^4}\frac{Q(2)^2\hbar^2}{\mu_{ab}}, \\
\label{gaan}
g_{aaN}v_{aa}'(r_{aa}) &= \frac{1}{N_a^2}\frac{r_{aa} }{R_0^4} \frac{Q(2)^2\hbar^2}{\mu_{ab}}
-\frac{4}{N_a (N_a -1)^2} \frac{1}{r_{aa}^3} \frac{Q(N_a)^2\hbar^2}{m_a},
\end{align}
where the following reduced mass is introduced
\begin{equation}
\mu_{ab} = \frac{N_a m_a m_b}{N_a m_a + m_b}.
\end{equation}
Taking into account that ${r'_0}^2=\frac{N_a-1}{2N_a}r_{aa}^2+R_0^2$, the intermediate positive quantities $r_{aa}$ and $R_0$ are linked by the relation
\begin{equation}
\begin{split}
\label{raarp0}
&v_{aa}(r_{aa})\left[ \frac{N_a-1}{2 N_a} \frac{r_{aa}}{R_0^4 v_{aa}'(r_{aa})}\frac{Q(2)^2}{\mu_{ab}}
-\frac{2}{N_a-1}\frac{1}{r_{aa}^3 v_{aa}'(r_{aa})} \frac{Q(N_a)^2}{m_a} \right] \\
&- v_{ab}(r_0') \frac{r_0'}{R_0^4 v_{ab}'(r_0')} \frac{Q(2)^2}{\mu_{ab}}= \frac{1}{N_a-1} \frac{1}{r_{aa}^2}\frac{Q(N_a)^2}{m_a}+\frac{1}{2 R_0^2}\frac{Q(2)^2}{\mu_{ab}} .
\end{split}
\end{equation}
This $N$-body system can only be bound if $g_{ab} > 0$. This is in agreement with (\ref{gabn}) which can only yield a positive value for $g_{abN}$, since $v_{ab}'(r_0')$ is expected to be negative. One can remark that the sign of $g_{aaN}$ is not well defined by (\ref{gaan}). Indeed, binding for the $N$-body system can occur with $g_{aa} \le 0$. Examples for such systems exist for long-range forces. This is the case for atoms in which the Coulomb repulsion between electrons does not destabilise the structure. This is also the case for the general nonrelativistic many-body harmonic oscillator Hamiltonian in which some stiffness coefficients can be null or negative \cite{cint21}. 

The case of a static source for the $N_a$ particles can be studied with $m_b \to \infty$, that is to say setting $\mu_{ab} = N_a m_a$. Within this limit, particle $b$ has no kinetic energy and rests at its position $\bm{r}_N$, which can be chosen as the origin of the coordinate system. It is worth mentioning that if the particle $b$ is identical to the particles $a$, the set of $N$ particles must be completely (anti)symmetrised, which implies that $p'_a=P_0$ and $r_{aa}=r_0'$ \cite{cimi22}. In these conditions, (\ref{gabn})-(\ref{raarp0}) reduces to (\ref{g})-(\ref{rho0}), as expected. If $N_a=1$, all quantities associated with the interactions between two particles $a$ disappear and (\ref{gabn})-(\ref{raarp0}) reduces to
\begin{align}
\label{gmu} 
&g_N= \frac{1}{\rho_0^2\, v(\rho_0)}\frac{Q(2)^2\hbar^2}{2 \mu_{ab}} \quad \textrm{with} \\
\label{rho0mu}
&2\, v(\rho_0) + \rho_0\, v'(\rho_0)=0,
\end{align}
where $v(r)=v_{ab}(r)$ and $\mu_{ab}=m_a m_b/(m_a + m_b)$. This result can be directly obtained from (35)-(37) in \cite{cimi22}, and it is in agreement with (\ref{g})-(\ref{rho0}) if $m_a=m_b=m$. If particle $b$ is a static source, then $\mu_{ab}=m_a$.  

For nonrelativistic systems, one can define $b_{ij}(x)$ such that $b_{ij}(x^2)= V_{ij}(x)$. If $d^2b_{aa}(x)/dx^2$ and $d^2b_{ab}(x)/dx^2$ are both concave (convex) functions for all positive values of $x$, an approximate ET energy is an upper (lower) bound of the genuine energy. If it is not the case, the variational character is not guaranteed. Again, upper (lower) bounds of the critical coupling constants are obtained if $E$ is an upper (lower) bound of the energy. As in the case of identical particles, only upper bounds of the energies can be computed with short-range potential wells if $g_{aa} \ge 0$. No variational character can be guaranteed for $E$ if $g_{aa} <0$.

The writing of (\ref{gabn})-(\ref{gaan}) is a little bit misleading because these equations do not give an unique pair of values for the two critical coupling constants, but instead a link between both ones. Indeed, bound states can appear for various couples of coupling constants. So, in fixing for instance the value of $g_{aa}$ ($g_{ab}$), the solution of (\ref{gabn})-(\ref{raarp0}) can yield the value of $g_{abN}$ ($g_{aaN}$), if any. So, the subscript $N$ should be associated with only $g_{aa}$ or $g_{ab}$ in (\ref{gabn})-(\ref{gaan}).

\begin{figure}[htb]
\includegraphics[height=5cm]{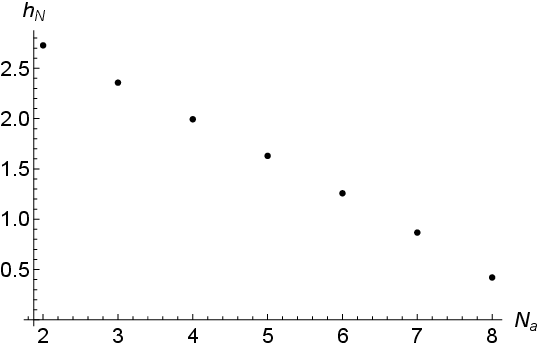}
\caption{Upper bounds of the critical coupling constant $h_N$ for the bosonic ground state computed from (\ref{tnp1a})-(\ref{tnp1c}) as a function of $N_a=N-1$ for $m=2$ and $g=1$. \label{fig:1}}
\end{figure}

\begin{figure}[htb]
\includegraphics[height=5cm]{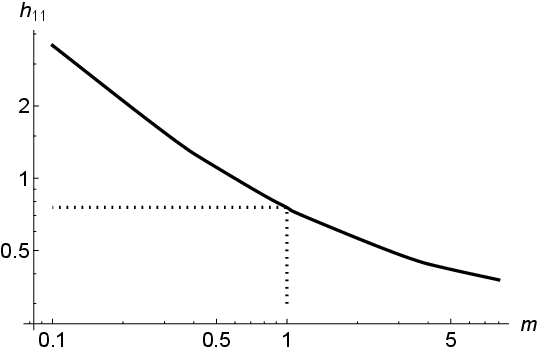}
\caption{Upper bounds of the critical coupling constant $h_{11}$ for the bosonic ground state computed from (\ref{tnp1a})-(\ref{tnp1c}) as a function of $m$ for $N_a = 10$ and $g=0.756$. \label{fig:2}}
\end{figure}

\begin{figure}[htb]
\includegraphics[height=5cm]{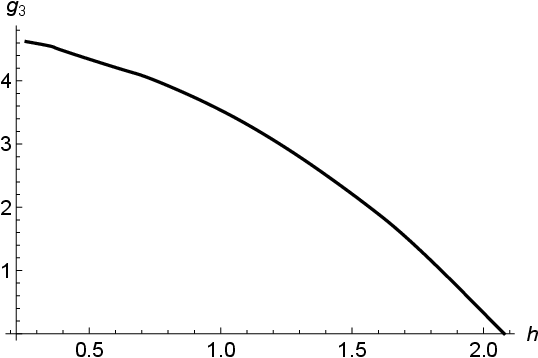}
\caption{Upper bounds of the critical coupling constant $g_{3}$ for the bosonic ground state computed from (\ref{tnp1a})-(\ref{tnp1c}) as a function of $h$ for two particles $a$ in a static source ($m = \infty$). \label{fig:3}}
\end{figure}

In order to illustrate what can be learned from (\ref{gabn})-(\ref{raarp0}), let us consider a simple situation in which $v_{aa}(r) = v_{ab}(r) = e^{-\mu r}$. Without loss of generality, one can set $\hbar=\mu=m_a=1$. In order to lighten the notation, we write $m_b=m$, $g_{aa}=g$ and $g_{ab}=h$ to follow notations used in \cite{rich94}. So, with these conventions, Hamiltonian~(\ref{HNp1}) becomes
\begin{equation}
\label{HNp1test}
H=\sum_{i=1}^{N_a} \frac{\bm p_i^2}{2} + \frac{\bm p_N^2}{2 m} -\sum_{i<j=2}^{N_a} g\,e^{|\bm{r}_i-\bm{r}_j|}-\sum_{i=1}^{N_a}h\,e^{|\bm{r}_i-\bm{r}_N|}.
\end{equation}
For this system, upper bounds for the eigenvalues and the critical coupling constants are provided by the ET if $g \ge 0$ ($h$ must be positive). For the bosonic ground state of (\ref{HNp1test}) with $D=3$, (\ref{gabn})-(\ref{raarp0}) become
\begin{align}
\label{tnp1a}
g_N\,e^{-r_{aa}} & = \frac{9}{N_a}\frac{1}{r_{aa}^3}-\frac{9}{4}\frac{N_a+m}{N_a^3 m}\frac{r_{aa}}{R_0^4}, \\
\label{tnp1b}
h_N\,e^{-r_0'} & = \frac{9}{4}\frac{N_a+m}{N_a^2 m}\frac{r_0'}{R_0^4} , \\
\label{tnp1c}
0 &= -\frac{(N_a-1)(N_a+m)}{N_a^2 m}\frac{r_{aa}}{R_0^4}+4 \frac{N_a-1}{r_{aa}^3}+2 \frac{N_a+m}{N_a m}\frac{r_0'}{R_0^4}-2\frac{N_a-1}{r_{aa}^2}-\frac{N_a+m}{N_a m}\frac{1}{R_0^2}.
\end{align}
Three examples of solutions for (\ref{tnp1a})-(\ref{tnp1c}) are given in Fig.~\ref{fig:1} to Fig.~\ref{fig:3}. 
In Fig.~\ref{fig:1}, the critical coupling constant $h_N$ is computed as a function of $N_a$ for $m=2$ and $g=1$. This constant decreases with the number of particles $a$, as expected. 
In Fig.~\ref{fig:2}, the critical coupling constant $h_{11}$ is computed from (\ref{tnp1a})-(\ref{tnp1c}) as a function of $m$ for $N_a = 10$ and $g=0.756$. This value of $g$ is the critical coupling constant, computed from (\ref{g})-(\ref{rho0}), to bind 11 identical particles with $m=1$. The graph shows that $h_{11}=0.756$ for $m=1$, as expected.
In Fig.~\ref{fig:3}, a system similar to the one presented in Fig.~1 of \cite{rich94} is studied. The critical coupling constant $g_{3}$ is computed as a function of $h$ for two particles $a$ in a static source ($m = \infty$). The value of the critical coupling constant to bind one particle $a$ in a static source can be computed from (\ref{gmu})-(\ref{rho0mu}) with $\mu_{ab}=m_a=1$, and it is $2.078$. So, two particles $a$ are just bind in the static source with $h=2.078$ if no repulsive interaction exist between them. The graph shows that $g_{3}=0$ for $h=2.078$, as expected. No binding is found below $h\approx 0.26$, since only solutions with negative values of $r_{aa}$ or $R_0$ are obtained.

\section{Summary}
\label{sec:sum}

Within the framework of the envelope theory \cite{sema13a,sema20}, we have computed upper bounds for critical coupling constants which allow the binding of some nonrelativistic quantum many-body systems in which particles interact only via two-body short-range central potential wells. In Sec.~\ref{sec:n}, a simple formula is obtained for $N$ identical particles. It is valid for bosons or fermions in dimension $D$ for any state of excitation. Two checks are performed for $N=2$ with $D=3$ and for $3 \le N  \le 100$ with $D=1$. It could be interesting to investigate numerically other systems with $N \ge 2$ to produce accurate values of critical coupling constants and test how far our upper bounds are from the exact ones. 

In Sec.~\ref{sec:npo}, relations between upper bounds for critical coupling constants have also been obtained for quantum systems with many identical particles plus a different one, for bosons or fermions in dimension $D$ for any state of excitation, although the situation is much more complex than for all identical particles. It is then possible to obtain information about the binding of the systems by solving a simple numerical set of three transcendental equations. Again, comparison with accurate numerical solutions of the many-body systems could be interesting. In principle, a similar analysis could be performed with the envelop theory for systems composed of two different sets of identical particles \cite{cimi22}, but the calculations should be much more involved.

The upper bounds obtained in this paper can yield information about the possible binding of complicated many-body systems. They can also be used to test numerical procedures. Only the nonrelativistic kinematics is considered here because it is the most interesting case and also because it leads to analytic formulas. Nevertheless, approximations for critical coupling constants could also be computed with the envelope theory for other Hamiltonians since this method can handle a general form of the kinetic energy. The semirelativistic form $T(p)=\sqrt{\bm p^2 c^2+m^2 c^4}$ is an obvious possibility. But more exotic forms exist, for instance in atomic physics with a nonparabolic dispersion relation \cite{arie92}, in hadronic physics with particle masses depending on the relative momentum \cite{szcz96}, in quantum mechanics with a minimal length \cite{brau99,ques10}, or in fractional quantum mechanics \cite{lask02}. Moreover, problems in $D$ dimensions can appear in various physical situations. In particular, $D = 2$ systems can be used as toy models for $D = 3$ systems \cite{tepe98} or are the natural framework for the physics of anyons \cite{khar05}. So, the possible domains of interest for the method developed here are numerous. Even if it is not possible to obtain analytical equations, numerical calculations can still easily be done whatever the number of particles. 

\appendix

\section{Hyperradial approximation}
\label{sec:hyper}

The bosonic ground state of Hamiltonian~(\ref{HNidD1}) ($D=1$) is accurately computed using a sophisticated hyperspherical harmonic expansion on Lagrange meshes in \cite{baye17}. At its lowest level, named as the $K=0$ limit in the paper, a quite reasonable approximation can be obtained by solving the following equation   
\begin{equation}
\label{K0}
\left( -\frac{d^2}{d\rho^2} +\frac{{\cal L}_0({\cal L}_0+1)}{\rho^2} -2\, m (E- V_{00}(\rho)) \right)\phi(\rho) = 0,
\end{equation}
with ${\cal L}_0=(N-4)/2$ and 
\begin{equation}
\label{V0K0}
V_{00}(\rho) = -\frac{N(N-1)}{2} g\, _1\textrm{F}_1\left( \frac{1}{2},\frac{N-1}{2},-2\,\mu^2\,\rho^2\right).
\end{equation}
This is illustrated in Table~\ref{tab:Gauss3}. One can remark that the quality of the approximation is deteriorating when the binding decreases in the system. It is tempting to use the procedure presented in~\cite{hult51} to compute an approximation of the critical coupling constant for the ground state of~(\ref{HNidD1}) from~(\ref{K0})-(\ref{V0K0}). But it is not clear that good results can be obtained since the $K=0$ limit does not seem reliable for situations in which the binding is vanishing. Nevertheless, (\ref{K0})-(\ref{V0K0}) should give good approximations of the ground state for coupling constants yielding an energy not close to zero. It is used in this spirit in Table~\ref{tab:Gauss2}. 

\begin{table}[H]
\begin{center}
\caption{Bosonic ground state for Hamiltonian~(\ref{HNidD1}). Accurate results from \cite{baye17} are compared with the ``$K=0$" approximation given by (\ref{K0})-(\ref{V0K0}). The parameters are taken from \cite{baye17}: $m=1/43.281307$, $\mu=\hbar=1$ and $g=10/\sqrt{\pi}$. 
\label{tab:Gauss3}}
\begin{tabular}{@{}ccccc}
\hline\hline
$N$ & 3 & 5 & 20 & 100 \\
\hline
\cite{baye17} & $-1.9325$ & $-9.2852$ & $-417.70$ & $-18552$ \\
(\ref{K0})-(\ref{V0K0}) & $-1.52$ & $-9.07$ & $-415.07$ & $-18539$ \\
\hline\hline
\end{tabular}
\end{center}
\end{table}

\begin{acknowledgments}
This work was supported by the IISN under Grant Number 4.45.10.08. 
\end{acknowledgments}

\end{document}